\documentclass[%
 aps,
 prl,%
 amsmath,amssymb,
reprint,%
superscriptaddress,showpacs]{revtex4-1}

\usepackage{color, graphicx}
\usepackage{dcolumn}
\usepackage{bm}
\usepackage{amssymb}
\usepackage{latexsym}
\usepackage{amsfonts}
\usepackage{amsmath}
\usepackage{multirow}
\usepackage{ifthen}

\begin{document}

\title{Perfect and broadband acoustic absorption by critical coupling}

\author{V. Romero-Garc\'ia}
\author{G. Theocharis}
\author{O. Richoux}
\author{A. Merkel}
\author{V. Tournat}
\author{V. Pagneux}
\affiliation{LUNAM Universit\'e, Universit\'e du Maine, CNRS, LAUM UMR 6613, Av. O. Messiaen, 72085 Le Mans, France}

\begin{abstract}
We experimentally and analytically report broadband and narrowband perfect absorption in two different acoustic waveguide-resonator geometries by the mechanism of critical coupling. In the first geometry the resonator (a Helmholtz resonator) is side-loaded to the waveguide and it has a moderate quality factor. In the second geometry the resonator (a viscoelastic porous plate) is in-line loaded and it contains two resonant modes with low quality factor. The interplay between the energy leakage of the resonant modes into the waveguide and the inherent losses of the system reveals a perfect and a broadband nearly perfect absorption. The results shown in this work can motivate relevant research for  the design of broadband perfect absorbers in other domains of wave physics. 
\end{abstract}


\maketitle

Structures composed of waveguides and resonators have been exploited for fundamental studies as well as for several applications in different branches of science and technology \cite{Bliokh08}. In these systems  the balance between the leakage rate of energy out of the resonator (resonator-waveguide coupling) and the inherent losses of the resonators has been shown of fundamental relevance for their transmission and reflection properties \cite{Slater50, Yariv00, Cai00}. When the leakage and the inherent losses are well balanced, the critical coupling condition is fulfilled, and then, a perfect destructive interference between the transmitted and the internal fields leads to maximum absorption at the resonance frequency \cite{Bliokh08, Xu00}. Therefore the critical coupling paved the way to produce perfect absorption (PA) of the incoming waves. In particular, the coherent perfect absorption \cite{Chong10}, being the time-reversed counterpart to laser emission \cite{Wan11}, makes use of the critical coupling mechanism to perfectly absorb the two port coherent excitation in a lossy medium. Recently, it has been shown that PA can be obtained by a single port excitation in a two port structure \cite{Piper14}. This is accomplished by the degenerate critical coupling condition, i.e., if two degenerated resonances are critically coupled and have opposite symmetry. 

PA is of particular interest for many applications such as energy conversion \cite{Law05, Tian07}, time-reversal technology \cite{Derode95, Fink97}, coherent perfect absorbers \cite{Chong10, Wan11} or soundproofing \cite{Mei12, Ma14} among others. In optics, several configurations using Bragg reflectors \cite{Tischler06}, Fabry-P\'erot cavities with metamaterial mirrors \cite{Deb10}, layered media with Kerr nonlinearity \cite{Reddy13} or graphene-based hyperbolic metamaterials \cite{Xiang14} have been recently used to obtain nearly PA of the incident radiation by critical coupling. Directional perfect absorbers have been also designed using deep subwavelength low-permittivity films \cite{Luk14}. Moreover PA has been obtained in plasmonic planar structures via the critical coupling of the surface plasmon modes \cite{Shin04}, while this has been also used for heat generation in plasmonic metamaterials \cite{Hao11}. In acoustics, coherent perfect absorption has been studied numerically by controlling the coherence of the input waves in a two port system showing a dynamically tuned absorption coefficient on the target material up to unity \cite{Wei14}.

\begin{figure}[h]
\includegraphics[width=8.5cm]{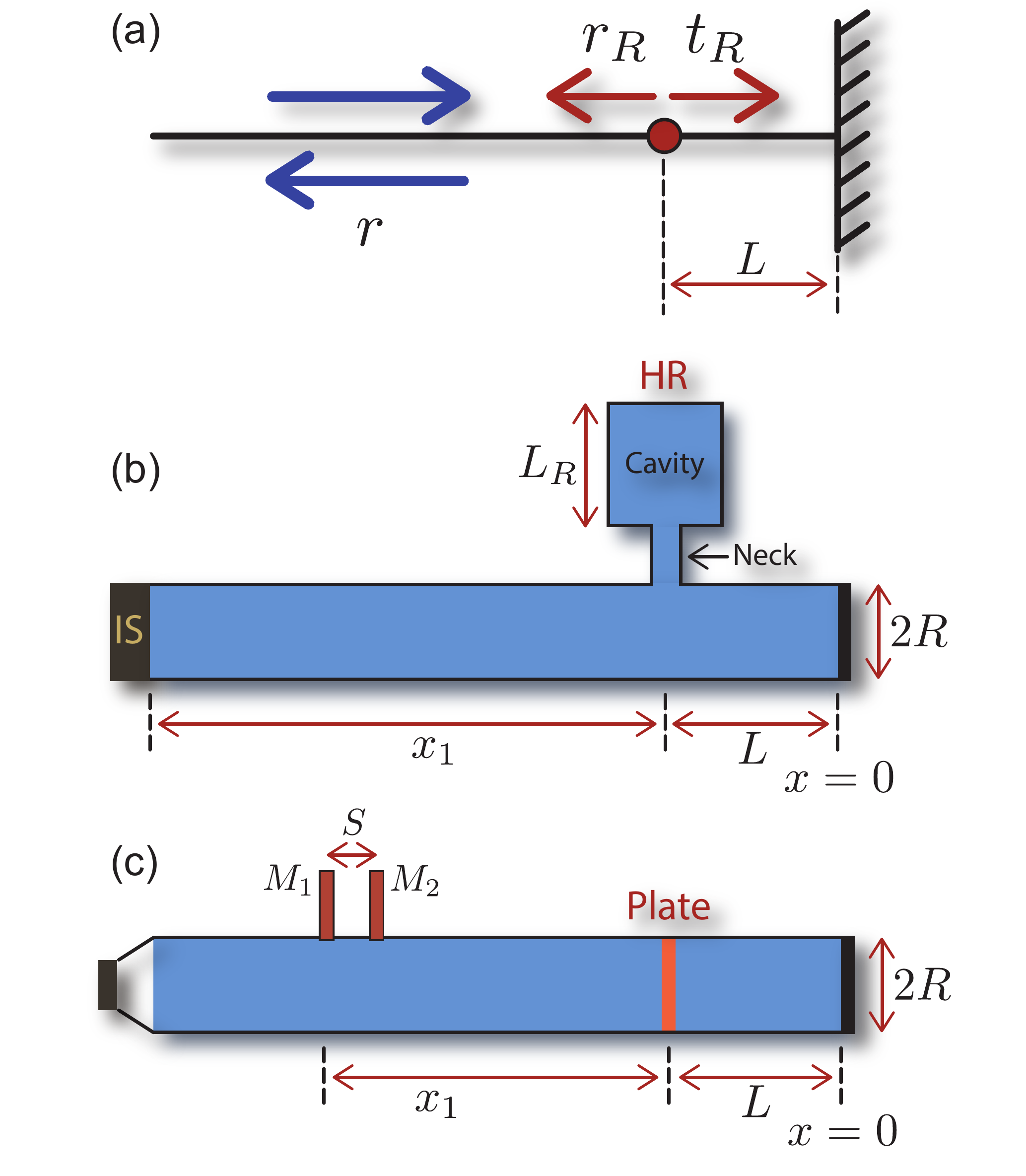}
\caption{(Color online) (a) Asymmetric Fabry-P\'erot resonator made of a resonant element (red point) and a rigid backing at distance $L$ from the resonator. (b) and (c) respectively show the side and in-line resonant coupling geometries backed by a cavity of length $L$. (b) experimental set-up using an impedance sensor ($IS$) as described in Ref. [\onlinecite{Theocharis14}]. The HR is composed of a neck of length $L_n=2$ cm with radius $R_n=1$ cm, a cavity with tunable length, $L_R$, and radius $R_R=2.15$ cm. The waveguide has a radius $R=2.5$ cm. (c) experimental set-up based on the ISO-10534-2 \cite{ISO}. The viscoelastic porous plate of radius $R_p=2.2$ cm is clamped in a tube of $R=2.2$ cm.}
 \label{fig:fig1} 
\end{figure}

In this study, we exploit the critical coupling to experimentally and theoretically show narrowband and broadband PA in two simple configurations: a side or an in-line resonant element coupled with a waveguide backed by a rigid end. The frequency range considered here is well below the first cut-off frequency of the higher propagative modes in the waveguide, therefore the problem is considered one-dimensional. Both configurations can be considered as an asymmetric Fabry-P\'erot cavity of length $L$ with two different mirrors, i.e., the resonant element and the rigid backing [as schematically shown in Fig. \ref{fig:fig1}(a)]. The absorption of this system can be expressed as $\alpha=1-|r|^2$, where $r$ is the complex reflection coefficient obtained from the standard three-medium layer Fresnel equation \cite{Born99}, 
\begin{equation}
r=r_R+\frac{t_R^2r_te^{\imath 2kL}}{1-r_Rr_te^{\imath 2kL}},
\label{eq:r}
\end{equation}
where the time harmonic convention is $e^{-\imath \omega t}$. $r_R$ and $r_t$ are the reflection coefficients of the resonant element and  of the termination, respectively (in our case, $r_t=1$), and $k=\omega/c$ is the wave number in the cavity of length $L$. Considering losses in this configuration, the PA is fulfilled when the reflection coefficient is zero, i.e., when the superposition of the multiple reflections in the cavity [second term in Eq.~(\ref{eq:r})] destructively interferes with the direct reflection from the resonant element (first term in Eq. (\ref{eq:r})). 

We use Helmholtz resonators (HRs) to build the side resonant coupling geometry [see Fig. \ref{fig:fig1}(b)], and viscoelastic porous plates to build the in-line resonant coupling geometry [see Fig.~\ref{fig:fig1}(c)]. We choose these two systems to analyze two different situations:  a weakly lossy single resonant element with moderate quality factor ($Q^{-1}\simeq0.1$, in the case of HRs) and a largely lossy double resonant element with low quality factor ($Q^{-1}\simeq1$, in the case of the viscoelastic porous plate). We show that PA can be achieved though the critical coupling of the resonant modes of the system. In the largely lossy case, we take advantage of the losses to overlap PA peaks with other nearly PA peaks and thus to obtain a broadband nearly PA. The two configurations have been experimentally analyzed using standard impedance tubes showing very good agreement with the theoretical predictions (see caption of Fig.~\ref{fig:fig1} for references and details of the set-ups).

To understand the origin of the PA in our systems, we analyze the reflection coefficient $r$ evaluated in the complex frequency plane, i.e., substituting $\omega=\omega_R+\imath \omega_I$ in $k$. This kind of map 
has been revealed extremely informative for understanding the PA process \cite{Luk14}. Without losses, one finds pairs of poles (at complex frequency $\omega_P$) and zeros (at complex frequency $\omega_Z=\omega^*_P$) of $r$ symmetrically distributed around the real frequency axis \cite{Pagneux13}. The zeros (poles) are in the positive (negative) half imaginary frequency plane \cite{Supplementary}. 
When the losses are introduced into the system, the zeros and the poles are in general down-shifted. In the case where the zero is on the real frequency axis ($\omega_Z=\omega_R+\imath0=2\pi f$), $r(\omega_Z)=0$ so the PA condition is fulfilled and the PA is obtained. 

\begin{figure*}[hbt]
\includegraphics[width=16.5cm]{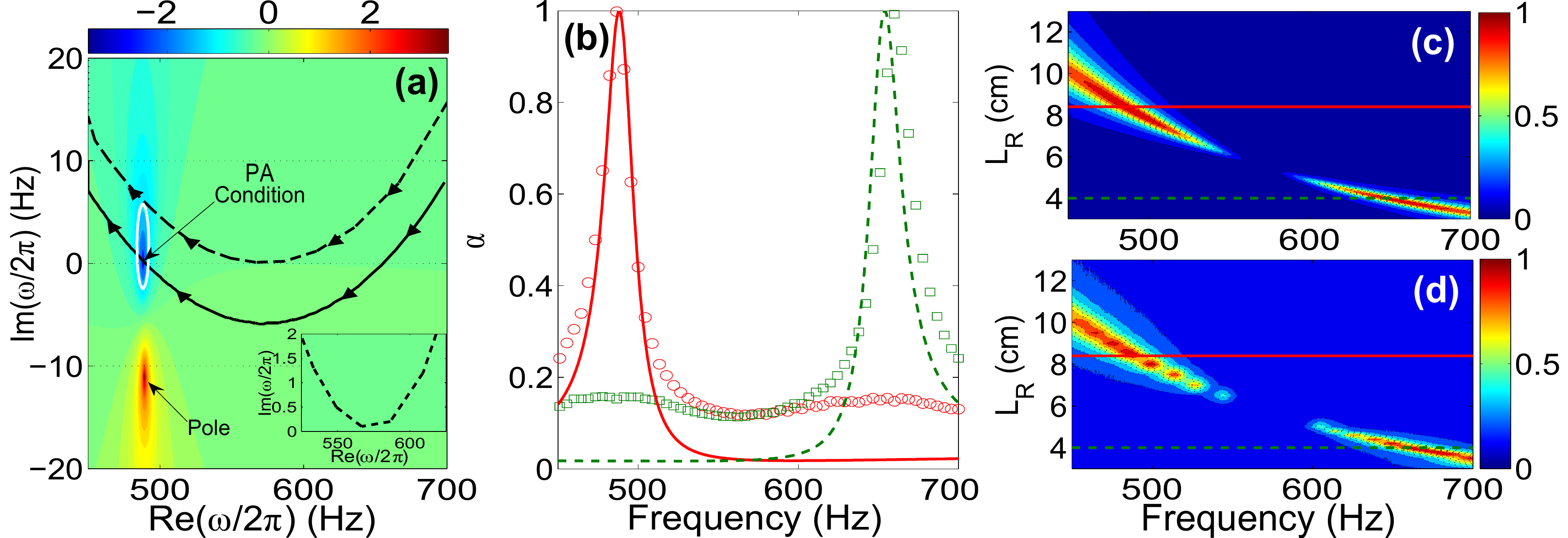}
\caption{(Color online) Side resonant geometry. (a) represents the complex frequency map of $\log{|r|}$ for the side resonant configuration with $L_R=8.3$ cm. Black dashed (Continuous) line represents the trajectory of the zero of $|r|$ for the lossless (lossy) case in the complex plane as $L_R$ is increased (sense of the increasing shown by arrows over the lines). White continuous line shows the isoline $\alpha=0.9$. Inset shows a zoom of the complex frequency map. (b) Green dashed (green open squares) and red continuous lines (red open circles)  show the theoretical (experimental) absorption coefficient, $\alpha=1-|r|^2$, for $L_R=3.9$ cm and $L_R=8.3$ cm respectively (horizontal lines in (c) and (d)). (c) and (d) show the theoretical and experimental maps respectively for the dependence of $\alpha$ on both the $L_R$ and the frequency, $f$.}
\label{fig:fig3}
\end{figure*}

We start by the analysis of the side resonant coupling geometry made of a HR with a variable cavity length, $L_R$, attached to a fixed backing cavity of $L=15$ cm [see Fig. \ref{fig:fig1}(b)]. This system presents viscothermal losses at the walls of the waveguide and of the resonator \cite{Theocharis14, Zwikker}. By changing $L_R$ from 0 to 15 cm, i.e., by changing the resonant frequency of the HR, we study the trajectory of the zero of the reflection coefficient in the complex frequency plane for both the lossless (black dashed line) and the lossy (black continuous line) cases in Fig. \ref{fig:fig3}(a). On one hand, as shown by the arrows over the trajectory, the zero moves to lower real frequencies as $L_R$ increases. On the other hand, the presence of the losses down-shifts the trajectory of the zero with respect to the lossless case 
and thus, the PA condition can be satisfied. In this case, there are two crossing points with the real frequency axis, i.e., two different configurations accomplishing the PA condition. These points correspond to $(L_R, f)=(8.3$ cm$, 484.5$ Hz$)$ and $(L_R, f)=(3.9$ cm$, 647$ Hz$)$. The blue and red areas in Figure \ref{fig:fig3}(a) respectively show the zero and the pole of the reflection coefficient [Eq.~(\ref{eq:r})] in the complex frequency plane for the configuration $(L_R, f)=(8.3$ cm$, 484.5$ Hz$)$. This case corresponds to a configuration at which the zero is on the real frequency axis, therefore the PA is physically observable. Similarly, a complex map for the configuration $(L_R, f)=(3.9$ cm$, 647$ Hz$)$ with the zero in the real frequency axis can also be obtained \cite{Supplementary}.

We now theoretically and experimentally evaluate the absorption coefficient $\alpha$ for these two configurations corresponding to the crossing points with the real frequency axis. We find analytically PA ($\alpha=1$) for the two above mentioned configurations at 484.5 Hz and 647 Hz as shown in Fig. \ref{fig:fig3}(b), in agreement with the crossing of the trajectories with the real frequency axis represented in Fig. \ref{fig:fig3}(a). Experiments show also very good agreement with the theoretical predictions, producing 100\% of absorption for these configurations at the corresponding frequencies of the PA condition. 

Finally, the theoretical and experimental dependencies of $\alpha$ on both $L_R$ and $f$ are plotted in Figs. \ref{fig:fig3}(c) and \ref{fig:fig3}(d) respectively. Several configurations, around the PA condition, are able to produce high absorption peaks in this system. In these configurations, the zero of the reflection coefficient is located close to the real frequency axis and influences the absorption coefficient that is physically observable. To characterize this high absorption, we define the region of nearly PA using the isoline $\alpha=0.9$ ($|r|^2=0.1$) around the zero in the complex frequency plane. If this isoline crosses the real frequency axis, a nearly PA peak with absorption $\alpha\geq0.9$ becomes observable \cite{Supplementary}.  
 This is translated in Figs.~\ref{fig:fig3}(c) and \ref{fig:fig3}(d) by the two regions with high absorption around the PA condition.

\begin{figure*}[hbt]
\includegraphics[width=16.5cm]{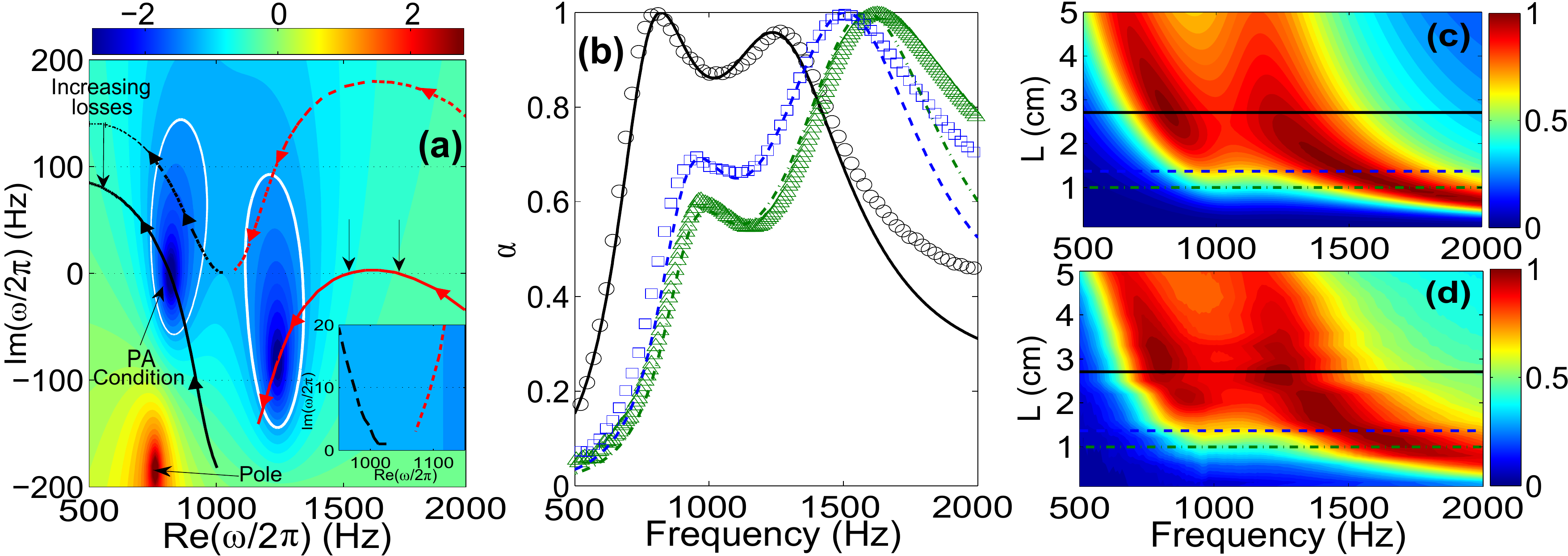}
\caption{(Color online) In-line resonant geometry. (a) represents the complex frequency map of $\log{|r|}$ for the in-line resonant configuration with $L=2.71$ cm. Black dashed (Continuous) and red dashed (continuous) lines represent the trajectory of the zero of $|r|$ for the first and second resonant modes for the lossless (lossy) case as $L_R$ is increased (sense of the increasing shown by arrows over the lines). White continuous line shows the isoline $\alpha=0.9$. Inset shows a zoom of the complex frequency map. In (b) green dot-dashed (green open triangles), blue dashed (open blue squares) and black continuous (black open circles) lines show  the analytical (experimental) absorption coefficients for the cases $L=1$ cm, $L=1.37$ cm and $L=2.71$ cm (horizontal lines in (c) and (d) respectively). (c) and (d) show the theoretical and experimental maps for the dependence of $\alpha$ on both the length of the cavity and the frequency.}
\label{fig:fig2}
\end{figure*}

The study now will be focused on the in-line resonant coupling geometry [Fig. \ref{fig:fig1}(c)]. This system presents larger inherent losses than the previous case, and also, it has two resonances in the studied frequency range. For that, a viscoelastic porous plate with losses coming from viscoelasticity and porosity is used as shown in Fig. \ref{fig:fig1}(c). The elastic properties are modelled by a Voigt rheological model, in which the Young's modulus is complex with a frequency dependent imaginary part, $E=E_0+\imath\eta\omega$ ($E_0=220$ kPa and $\eta=7$ Pa s). Due to the porosity, the mass density is also complex with the form $\rho=\rho_p-\imath\chi/\sqrt{\omega}$ ($\rho_p=28$ kg/m$^3$ and $\chi=350$ kg m$^3$ s$^{1/2}$). In both cases, $\eta$ and $\chi$ are extracted from the experimental results \cite{Supplementary}. Recently, a similar structure has been analyzed \cite{Ma14} showing that a small amount of losses can lead  to the design of subwavelength impedance matched systems. In our work, the inherent losses are substantially larger and the system contains two resonant modes. Therefore, the effects of each mode can overlap to produce a broadband nearly PA. 

In Fig. \ref{fig:fig2}(a), we analyze the trajectory of the zeros of Eq.~(\ref{eq:r}) for both cases, the lossless and the lossy, as we change the cavity length $L$ from 0 to 13 cm. As shown by the arrows over the trajectories, the two zeros move to lower real frequencies as the cavity length $L$ increases. 
When the inherent losses are taken into account, the trajectory of the zero of the first mode crosses the real frequency axis corresponding to the configuration $(L,f)=(2.71~\text{cm}, \quad 820~\text{Hz})$. Moreover, the trajectory of the zero of the second resonance crosses the real frequency axis in two points, corresponding to the configurations $(L,f)=(1$ cm$, 1703$ Hz$)$ and $(L,f)=(1.37$ cm$, 1508$ Hz$)$. The color map in Fig.~\ref{fig:fig2}(a) depicts the evaluation in the complex frequency plane of Eq.~(\ref{eq:r}) for the configuration $(L,f)=(2.71$ cm$, 820$ Hz$)$. The zero of the reflection coefficient of the first resonant mode is on the real frequency axis, while for the second resonant mode it is far away from the real frequency axis. However, the isoline $\alpha=0.9$ around the zero of the second resonant mode crosses the real frequency axis, denoting a substantial increase of the absorption in this frequency range. Figure \ref{fig:fig2}(b) represents the absorption coefficient evaluated for the configurations having the zero of the reflection coefficient in the real frequency axis, i.e., with the PA condition activated. For the configuration with $L=2.71$ cm (black line), the absorption peak due to the first mode is a signature of the PA (experimentally 100\% of absorption), while the second one produces a nearly PA peak explained by the crossing of the isoline $\alpha=0.9$ with the real frequency axis [see Fig. \ref{fig:fig2}(a)]. Therefore the zero produced by the second mode, although being far away from the real frequency axis, influences the absorption coefficient and overlaps with the PA peak of the first mode producing a broadband absorption peak with nearly PA. The other configurations, $L=1$ cm and $L=1.37$ cm, clearly show the PA peak produced by the critical coupling of the second resonant mode [see arrows in Fig. \ref{fig:fig2}(a)].

Finally, in Figs. \ref{fig:fig2}(c) and \ref{fig:fig2}(d), we respectively plot the theoretical and experimental dependency of $\alpha$ on both $L$ and $f$. Two zones, corresponding to the first and second resonant modes, with very high absorption around the PA configurations are visible. PA peaks appear only for three particular configurations [horizontal lines, corresponding to the results shown in Fig. \ref{fig:fig2}(b)], but the areas around them still present very high absorption. It is worth noting here that, in contrast to the case of the HRs with moderate quality factor, the viscoelastic porous plates have a low quality factor. Therefore, even for the configurations with the zero of the reflection coefficient out of the real frequency axis, the isoline $\alpha=0.9$ is wide enough to cross the real frequency axis and generate broad areas with nearly PA. 

In conclusion, we have exploited the critical coupling to experimentally and analytically show narrowband and broadband PA of acoustic waves in two waveguide-resonator geometries: the side and the in-line resonant coupling geometries. 
The complex frequency maps of the reflection coefficient and the trajectories of its zeros in the complex frequency plane have been revealed very insightful. The presence of losses produces a down-shift of the lossless trajectory of the zero inducing its crossing with the real frequency axis. The PA occurs when the zero of the reflection coefficient is on the real frequency axis. Nearly PA becomes observable when the isoline $\alpha=0.9$ crosses the real frequency axis. We have shown here that, making use of the losses of the system, one can overlap PA and nearly PA peaks to produce broadband PA absorption. This could allow challenging applications and 
advances in different domains of wave physics.

\begin{acknowledgments}
This work has been funded by the Metaudible project ANR-13-BS09-0003, co-funded by ANR and FRAE. Authors acknowledge the help of D. Parmentier in the experimental set-ups. GT acknowledges financial support from the FP7-People-2013-CIG grant, Project 618322 ComGranSol.
\end{acknowledgments}


%

\end{document}